\documentclass[aps,prd,superscriptaddress,10pt,onecolumn]{revtex4}
\usepackage{amssymb}
\usepackage{amsmath}
\usepackage{amsfonts}
\usepackage{graphicx, float}
\usepackage{graphicx, epsfig}
\usepackage{color}
\usepackage{enumerate}

\newcommand{\beq}{\begin{equation}}
\newcommand{\eeq}{\end{equation}}
\newcommand{\bea}{\begin{eqnarray}}
\newcommand{\eea}{\end{eqnarray}}

\begin{document}

\title{Instability and phase transitions of a rotating black hole\\ in the presence of perfect fluid dark
matter}

\author{Seyed Hossein Hendi}
\email{hendi@shirazu.ac.ir} \affiliation{Physics Department and
Biruni Observatory, College of Sciences, Shiraz University, Shiraz
71454, Iran} \affiliation{Research Institute for Astronomy and
Astrophysics of Maragha (RIAAM), P.O. Box 55134-441, Maragha,
Iran}

\author{Azadeh Nemati}
\affiliation{Physics Department and Biruni Observatory, College of
Sciences, Shiraz University, Shiraz 71454, Iran}

\author{Kai Lin}
\affiliation{Institute of Geophysics and Geomatics, China
University of Geosciences, Wuhan, Hubei 430074, China}

\author{Mubasher Jamil}
\email{mjamil@zjut.edu.cn} \affiliation{Institute for Theoretical
Physics and Cosmology, Zhejiang University of Technology,
Hangzhou, 310023 China} \affiliation{United Centre for
Gravitational Wave Research, Zhejiang University of Technology,
Hangzhou 310023, China} \affiliation{School of Natural Sciences,
National University of Sciences and Technology, H-12, Islamabad
44000, Pakistan}

\begin{abstract}
In this paper, we study the thermodynamic features of a rotating
black hole surrounded by perfect fluid dark matter. We analyze the
critical behavior of the black hole by considering the known
relationship between pressure and cosmological constant. We show
that the black hole admits a first order phase transition and,
both rotation and perfect fluid dark matter parameters have a
significant impact on the critical quantities. We also introduce a
new ad hoc pressure related to the perfect fluid dark matter and
find a first order van der Waals like phase transition. In
addition, using the sixth order WKB method, we investigate the
massless scalar quasinormal modes (QNMs) for the static
spherically symmetric black hole surrounded by dark matter. Using
the finite difference scheme, the dynamical evolution of the QNMs
is also discussed for different values of angular momentum and
overtone parameters.
\end{abstract}

\maketitle

\section{Introduction}

Black hole thermodynamics continues to be a promising topic of
gravitational physics as it is one of the possible routes towards
quantum gravity. The idea of relating black holes and ordinary
thermodynamics was notably founded by Bekenstein and Hawking and
subsequently carried on by other researchers. More recently the
idea of thermodynamics of anti-de Sitter (AdS) black holes has got
unusual attention by the discovery of gauge-gravity (AdS/CFT)
correspondence \cite{mal}. Hawking and Page found the existence of
a phase transition between the stable Schwarzschild black hole and
the pure radiation or gas \cite{page}. Later it was found that
there exists a small-large black hole phase transition for charged
or rotating AdS black holes \cite{cha1,cha2}. This phase
transition is also identified with the liquid-gas phase transition
of the van der Waals fluid \cite{mann}. We can work in the
extended phase space thermodynamics and use the key identification
between the cosmological constant and the pressure by $P\sim
\Lambda$, and its conjugate variable as the thermodynamic volume
\cite{PV1,PV2}.

In literature, the critical phenomenon and the phase transition of
several AdS black holes in various gravitational setups have
already been explored. 
The first order phase transition for five
dimensional charged AdS black holes was investigated in
\cite{wei}. Similar critical behavior was also observed for the
AdS black hole in massive gravity \cite{sf}.  
A more general treatment of phase transitions for extremal black
holes was proposed in \cite{p} without considering any specific
black hole. Wei and Liu proposed an interesting connection between
the impact parameter of the photon orbits and the thermodynamic
phase transitions of charged AdS black holes \cite{wei2}. They
suggested that the changes of the photon sphere radius and the
minimum impact parameter can serve as order parameters for a
small-large black hole phase transition.

Thermodynamics of Kerr-AdS black hole in four and higher
dimensions is discussed in Ref. \cite{Altamirano:2014tva}. It is
shown that in the canonical ensemble (fixed angular momentum),
four dimensional rotating AdS black hole behaves qualitatively
similar to the charged AdS case with fixed angular momentum
replacing fixed charge. The critical point for every fixed angular
momentum can be determined numerically. They demonstrated that in
the regime of slow rotation, a van der Waals type phase transition
takes place. In \cite{Czinner:2017tjq}, the R\'{e}nyi approach is
used to study thermodynamics of Kerr black hole. In this approach,
all the thermodynamic quantities are obtained using
R$\acute{e}$nyi entropy rather than Bekenstein-Hawking entropy. It
is shown that the characteristic swallow-tail behavior is observed
for the definite R\'{e}nyi entropy parameter, which is
corresponding to a small/large black hole phase transition
analogous to the picture of rotating black holes in AdS space.

It is worth mentioning the possibility of studying the extended
phase space from the viewpoint of AdS/CFT correspondence. In this
mechanism, the AdS radius relates to the number of colors in the
dual gauge theory. So, variable cosmological constant is
equivalent to variable number of colors \cite{Kubiznak:2016qmn}.
Moreover, the thermodynamic volume may be interpreted as
associated chemical potential for the color. It is known that
changing the sign of chemical potential is an indication of
quantum effects. This subject has been studied in
\cite{Maity:2015ida} for Kerr-AdS black holes in four and five
dimensions. It is shown that the sign of chemical potential is
changed above the Hawking-Page transition temperature which is
physically dual to a confinement-deconfinement transition of the
boundary gauge theory.

On the other hand, it is interesting to study the effect of
surroundings on the thermodynamics of rotating black hole. The
Kerr black hole in the presence of electromagnetic field is
considered in Ref. \cite{Liao:2016mum}. The authors explored the
parameter condition for superconducting phase transition and
obtained an appropriate value for the ratio of mass squared over
angular momentum so that Meissner effect occurs. In other words,
conditions for existence/non-existence of Meissner effect phase
transition is discussed. Phase transition for the Kerr-Newman-AdS
black hole in quintessence matter as a model of dark energy is
discussed in \cite{Jafarzade:2017wtc}. Extension of the model
including nonlinear magnetic charge is given in
\cite{Ndongmo:2019ywh}. In this paper, we are interested to
explore the effect of dark matter on the rotating black holes.

It is suggested that about $27$ \% of our universe is made from
the invisible dark matter. One of the main evidence for this idea
is the galaxy's flat rotation curve \cite{Rubin}. The dependence
of rotation velocity on the distance from the center of spiral
galaxies has an asymptotically flat character in contrast to the
expectations from Newton's law, while the strong dominated
gravitational field due to the presence of dark matter in far
distances can explain this observation. For an explanation of this
dark halo contribution, Kiselev presented a new class of solutions
of the Einstein equation, by applying the perfect fluid relations,
and introduced a new logarithmic term which explains the observed
asymptotic behavior at large distances \cite{kiselev}. Li \& Yang
proposed a model of black hole immersed in dark matter halo in the
presence of dark energy modeled as inhomogeneous phantom field
\cite{yang}. They extended the Kiselev solution to the exact
static spherically solution consistent with the Schwarzschild-AdS
and Reissner-Nordstr\"{o}m metrics. Later Xu et al generalized
these solutions to the Kerr dS/AdS black holes surrounded by the
perfect fluid dark matter using the Janis-Newman algorithm
\cite{xu1}. In literature, astrophysical aspects of this black
hole have already been studied such as shadow images and the
geodetic precession frequency \cite{sum,r}.

In this paper, we are interested to investigate how the presence
of perfect fluid dark matter affect the thermodynamical/dynamical
aspects of black holes. We examine dynamical stability with
quasi-normal modes (QNMs) along with thermal stability and
possible phase transition.

Regarding different black hole solutions of gravitating systems,
the examination of instability conditions is a strong tool to veto
some models. In order to obtain the stability criteria and
investigate the stability of a black hole, one should examine its
response to dynamic and thermodynamic perturbations. On the one
hand, the behavior of the heat capacity is one of the powerful
tools to analyze thermal stability. It is shown that in the
canonical ensemble (fixed charged) the positivity of heat capacity
can guarantee (local) thermal stability. On the other hand, the
QNMs \cite{RW} can reflect the behavior of black holes under
dynamic perturbations \cite{QNM1,QNM2,QNM3,QNM4}. After the
detection of the gravitational radiation of compact binary mergers
by LIGO and VIRGO observatories \cite{GW1,GW2,GW3}, investigation
of QNMs attracted much attention. This is due to the fact that the
spectrum of gravitational QNMs perturbations can be traced by
gravitational wave detectors \cite{PRL}. In this paper, we
restrict ourselves to the case of scalar perturbation.

The plan of the paper is as follows: In section II, we discuss the
thermodynamical properties of the black hole under consideration.
The phase transitions analysis is performed in section III. We
then investigate QNMs for the static black hole in section IV, and
finally, conclude in section V.

\section{Rotating perfect fluid dark matter black hole and its thermodynamics}

We consider a black hole with rotation immersed in a perfect fluid
background \cite{sum}
\begin{equation*}
ds^{2}=-\frac{\Delta _{r}}{\rho ^{2}\Sigma }(dt-a\sin ^{2}\theta d\phi )^{2}+%
\frac{\Delta _{\theta }\sin ^{2}\theta }{\rho ^{2}\Sigma }%
(adt-(r^{2}+a^{2})d\phi )^{2}+\frac{\Sigma }{\Delta _{r}}dr^{2}+\frac{\Sigma
}{\Delta _{\theta }}d\theta ^{2},
\end{equation*}%
where
\begin{eqnarray}
\Delta _{r} &=&r^{2}-2Mr+a^{2}-\frac{\Lambda }{3}r^{2}(r^{2}+a^{2})+\alpha
r\ln (r/|\alpha |),  \nonumber \\
\Delta _{\theta } &=&1+\frac{\Lambda }{3}a^{2}\cos ^{2}\theta ,  \nonumber \\
\rho &=&1+\frac{\Lambda }{3}a^{2},  \nonumber \\
\Sigma &=&r^{2}+a^{2}\cos ^{2}\theta .
\end{eqnarray}
The logarithmic term in the metric is responsible for presence of
dark matter and the intensity of the PFDM is presented by the
parameter $ \alpha $. We see that this solution reduces to the
Kerr-AdS metric as we set $ \alpha=0 $.

The mass of the black hole is determined by the condition $\Delta
_{r}(r_{+})=0$, so
\begin{equation}
M=\frac{1}{2}\Big[r_{+}+\frac{a^{2}}{r_{+}}-\frac{\Lambda }{3}%
r_{+}(r_{+}^{2}+a^{2})+\alpha \log \Big(\frac{r_{+}}{|\alpha
|}\Big)\Big].
\end{equation}

\begin{figure}[h!]
\begin{center}
$%
\begin{array}{cc}
\includegraphics[width=80 mm]{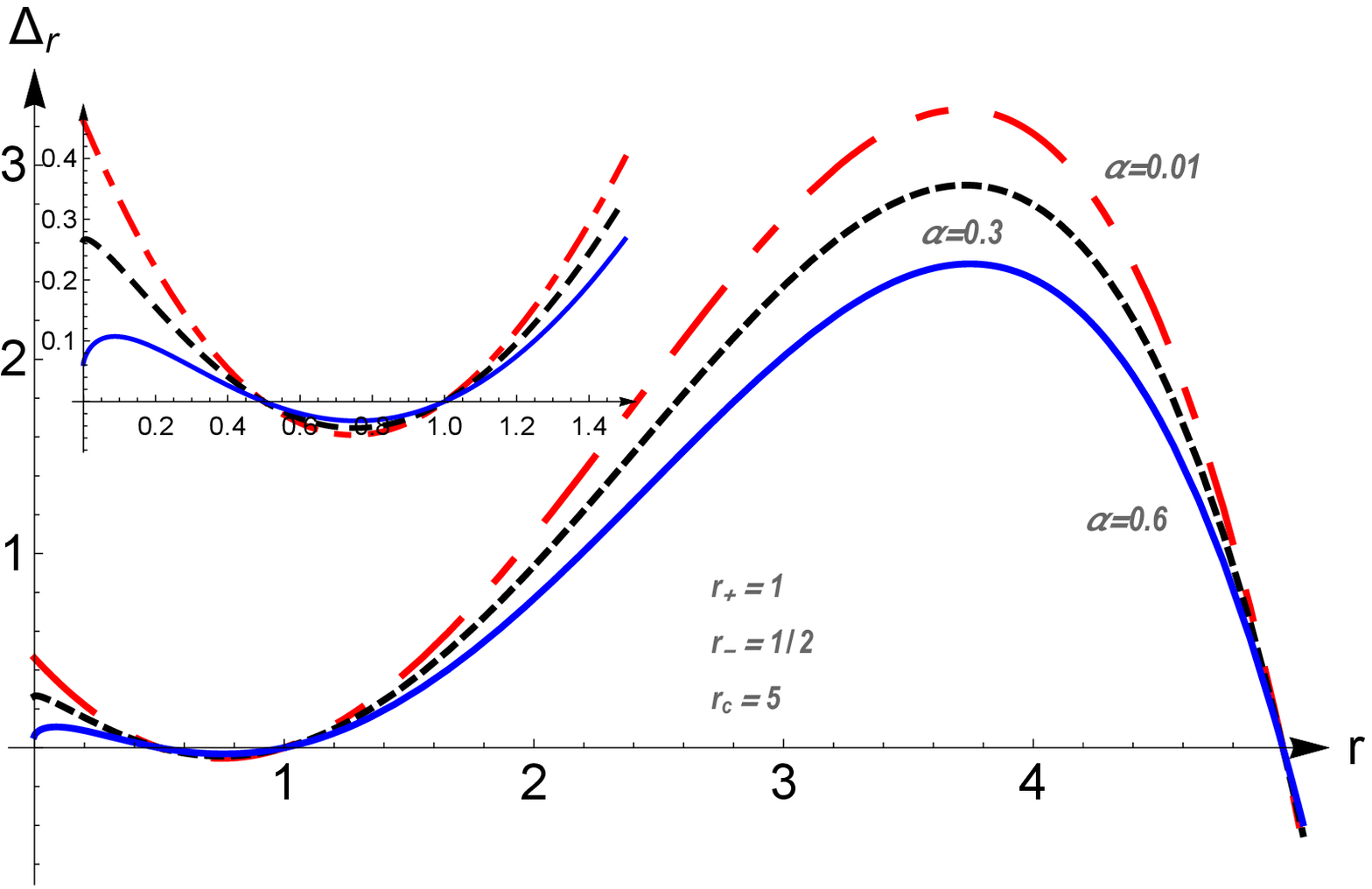} & \includegraphics[width=80 mm]{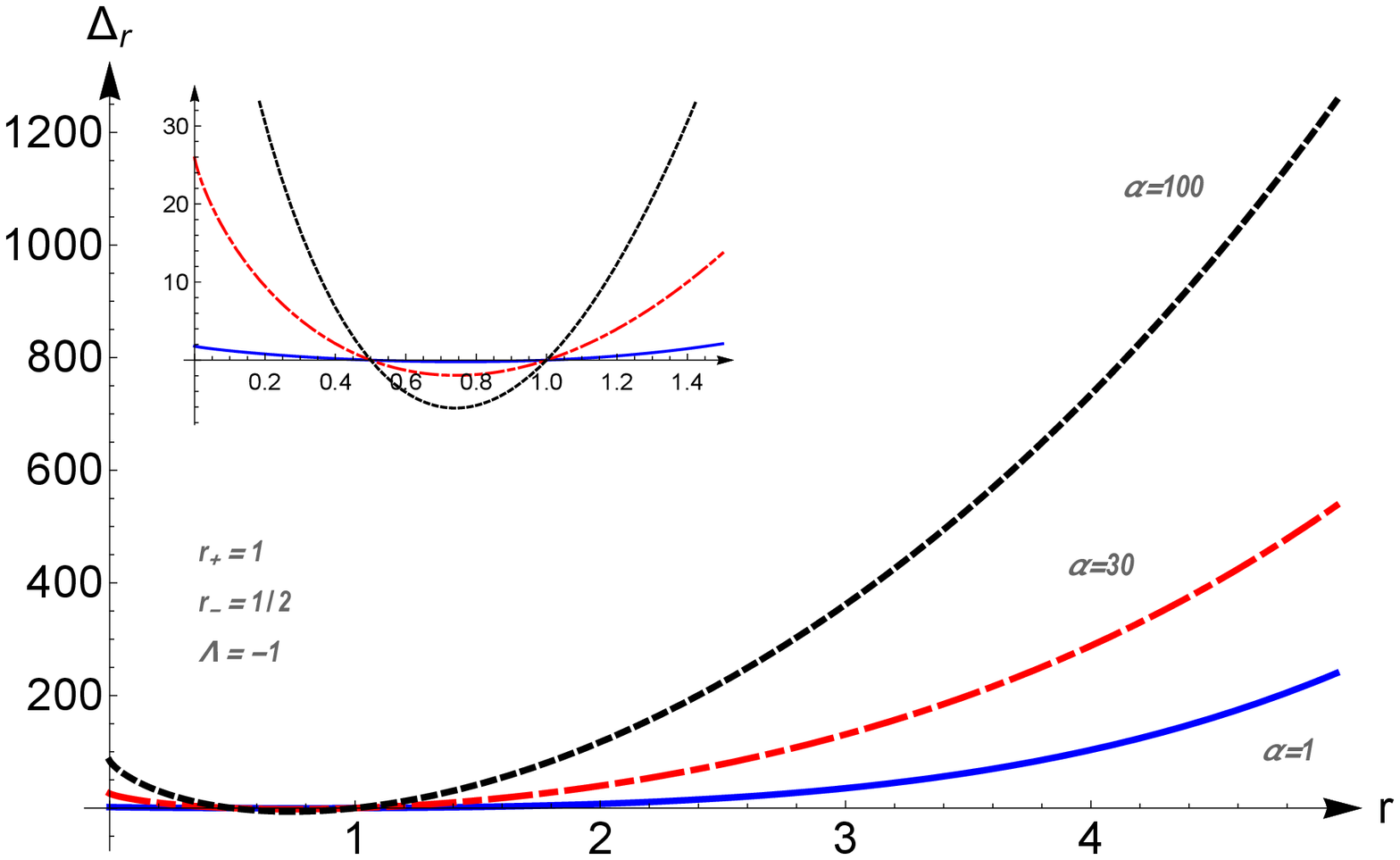}
\end{array}%
$%
\end{center}
\caption{$\Delta_r$ versus $r$ with different $\protect%
\alpha$, where $r_+=1$, $r_-=1/2$: (left panel: dS solutions with
$r_c=5$ and right panel: AdS solutions with $\Lambda=-1$).}
\label{FIG1}
\end{figure}

Regarding asymptotically dS solutions (left panel of Fig.
\ref{FIG1}), the event horizon $r_+$, inner horizon $r_-$ and
cosmological horizon $r_c$ of black hole satisfy
$\Delta_r(r_+)=\Delta_r(r_-)=\Delta_r(r_c)=0$, so we can use the
position of horizons to replace the parameters $M$, $a$ and
$\Lambda$. In addition, the behavior of asymptotically AdS
solutions is shown in the right panel of Fig. \ref{FIG1}. Briefly,
the relation between $\Delta_r $ and $r$ with different $\alpha$
is shown in Fig. \ref{FIG1}.

The area of event horizon of the black hole is given by
\begin{equation}
A=4\pi \frac{(r_{+}^{2}+a^{2})}{\rho }.
\end{equation}%
\newline
Therefore the Bekenstein-Hawking entropy relation becomes
\begin{equation}
S=\frac{A}{4}=\pi \frac{(r_{+}^{2}+a^{2})}{\rho },
\end{equation}%
and Hawking temperature associated with the surface gravity of the event
horizon $r_{+}$ is determined as:
\begin{equation}  \label{temp}
T_{h}=\frac{r_{+}}{4\pi \rho (r_{+}^{2}+a^{2})}\Big[1-\frac{a^{2}}{r_{+}^{2}}%
-\frac{\Lambda }{3}(3 r_{+}^{2}+a^{2})+\frac{\alpha }{r_{+}}\Big].
\end{equation}



\section{Phase transitions}

In this section, we are going to examine possible phase
transition. Following the traditional $PV$ criticality in the
extended phase space, we can regard the cosmological constant as a
dynamical pressure, $\Lambda =- 8\pi P$. Considering Eq.
(\ref{temp}), it is straightforward to obtain the following
equation of state
\begin{equation}
P=\frac{3}{8\pi a^2}\left[\frac{4 \pi T (r_{+}^2+a^2) -(r_{+} +
\alpha)+\frac{a^2}{r_{+}}}{4 \pi T  (r_{+}^2+a^2) +r_{+} +\frac{3
r_{+}^3}{a^2}}  \right]. \label{P1}
\end{equation}%
Here, we can identify the specific volume $v$ as $v=2r_{+}$ in the
geometric units. Therefore, as it is usual, we use the event
horizon radius instead of the specific volume in order to analyze
the criticality. Due to the fact that the critical point in the
isothermal $P-r_{+}$ is an inflection point, we can find the
critical quantities with the following equations
\begin{equation}
\left( \frac{\partial P}{\partial r_{+}}\right) _{T}=0,\text{ \ \ \ \ \ \ }%
\left( \frac{\partial ^{2}P}{\partial r_{+}^{2}}\right) _{T}=0.
\label{dp-ddp}
\end{equation}

\begin{figure}[h!]
\begin{center}
$%
\begin{array}{cc}
\includegraphics[width=80 mm]{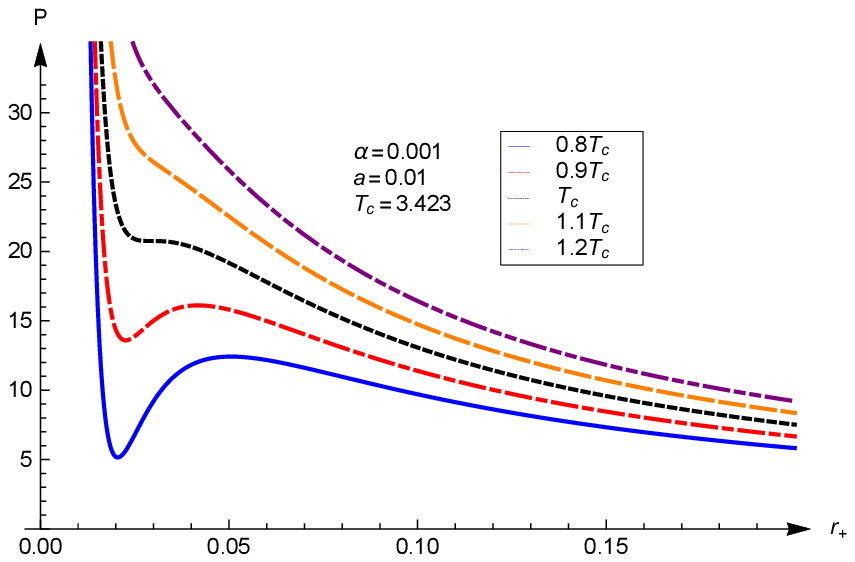} & %
\includegraphics[width=80 mm]{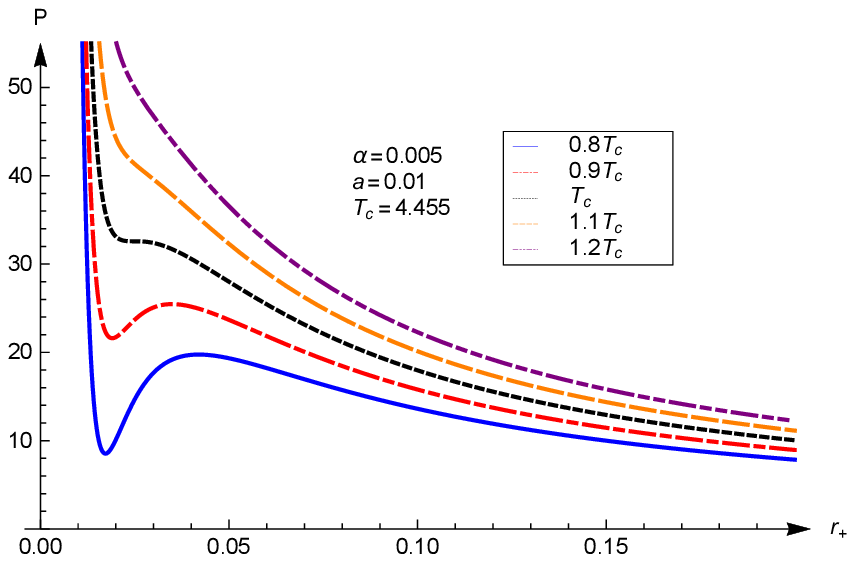}%
\end{array}%
$%
\end{center}
\caption{Pressure versus $r_{+}$ for $a=0.01$ and
$\protect\alpha=0.001$ (left panel), and $\protect\alpha=0.005$
(right panel). } \label{FigPV-Lambda}
\end{figure}

Since the analytical calculation of the critical quantities is not
a trivial task, we use the numerical analysis. Using Eq.
(\ref{dp-ddp}), the critical point can be found numerically.
Regarding the functional form of the pressure, one finds that its
denominator can be vanished for some values of the event horizon
radius (we called largest ones as $r_{+d}$). Investigating such
divergencies with more details, we find that the pressure is negative for $%
r_{+}<r_{+d}$, and therefore, we study the positive pressure region, $%
r_{+}>r_{+d}$. Looking at Fig. \ref{FigPV-Lambda}, we find the van
der Waals like behavior for the solution, and therefore, a first
order phase transition has been occurred for $T<T_{c}$. The
critical values for the mentioned black hole solutions are
addressed in table I. In this table, we focus on the effects of
$a$ and $\alpha$. It is observed that increasing the rotation
parameter (decreasing $\alpha$) leads to increasing the critical
horizon radius and decreasing the critical temperature and
pressure. In other words, the effect of $a$ and $ \alpha $ on the
critical quantities is opposed to each other. It means that the
criticality is easier to see for highly rotating black holes with
vanishing $\alpha$. In Ref. \cite{Altamirano:2014tva}, by
approximating the equation of state in the regime of slow
rotation, one can find that decreasing the rotation parameter
leads to increasing the critical pressure and temperature
(decreasing the critical radius) which is consistent with our
results in the table I. Influence of perfect fluid dark matter on
the thermodynamic behavior of Reissner-Nordstr\"{o}m-AdS black
hole is investigated in Ref. \cite{Xu:2016ylr}. In the mentioned
paper, phase transition is discussed in charge squared-electric
potential plane and a resemblance to the van der Waals system is
found.

\begin{center}
\begin{tabular}{|c|c|c|}
\hline
\begin{tabular}{cccc}
\hline\hline
\hspace{0.4cm}$a$\hspace{0.4cm} & \hspace{0.4cm}$r_{c}$ \hspace{0.4cm} &
\hspace{0.4cm} $T_{c}$\hspace{0.4cm} & \hspace{0.4cm} $P_{c}$\hspace{0.4cm}
\\ \hline\hline
$0.01$ & $0.0295$ & $3.423$ & $20.75$ \\ \hline
$0.05$ & $0.1529$ & $0.6501$ & $0.7578$ \\ \hline
$0.10$ & $0.3073$ & $0.3229$ & $0.1873$ \\ \hline
$0.20$ & $0.6160$ & $0.1610$ & $0.0465$ \\ \hline
\end{tabular}
& \ \ \ \ \ \  &
\begin{tabular}{cccc}
\hline\hline
\hspace{0.4cm}$\alpha $\hspace{0.4cm} & \hspace{0.4cm}$r_{c}$ \hspace{0.4cm}
& \hspace{0.4cm} $T_{c}$\hspace{0.4cm} & \hspace{0.4cm} $P_{c}$\hspace{0.4cm}
\\ \hline\hline
$0.005$ & $0.02482$ & $4.455$ & $32.58$ \\ \hline
$0.010$ & $0.02052$ & $6.186$ & $55.38$ \\ \hline
$0.030$ & $0.01268$ & $19.56$ & $253.6$ \\ \hline
$0.050$ & $0.01000$ & $55.70$ & $596.8$ \\ \hline
\end{tabular}
\\ \hline
\end{tabular}
\\[0pt]
\textbf{Table I:} Left table: critical values for $\alpha =0.001$
with different rotation
parameter.\\[0pt]
Right table: critical values for $a=0.01$ with different $\alpha $.
\end{center}


To get more information about the phase transition, we study
thermodynamic quantities such as heat capacity and Gibbs free
energy. Using the standard definition, they are as follows

\begin{equation}
C_{P}=  T \frac{\partial S}{\partial T} \bigg\vert_{\Lambda,a
,\alpha}   =\frac   {6 \pi (a^2+r_{+}^2)  \left(  8 \pi P
(r_{+}^2+\frac{a^2}{3})+1+\frac{\alpha}{r_{+}}-\frac{a^2}{r_{+}^2}
\right)   }  {(8 \pi P a^2-3)  \left (-8 \pi P (r_{+}^2+\frac{8
a^2}{3}+\frac{a^4}{3 r_{+}^2})+1+\frac{2
\alpha}{r_{+}}-\frac{a^2}{r_{+}^2} (4+\frac{a^2}{r_{+}^2})
\right)   },
\end{equation}
\begin{equation}
G=M-T S=- \frac {r_{+} \left(1-\frac{a^2}{r_{+}^2}+ \frac{8}{3}
\pi P (a^2+3 r_{+}^2) +\frac{\alpha}{r_{+}} \right)}{ 4 \left(
1+\frac{\Lambda a^2}{3} \right) \left(  1-\frac{8}{3} \pi P a^2
\right)}+\frac{1}{2} \left( r_{+}+\frac{a^2}{r_{+}} +\frac{8}{3}
\pi P r_{+} (a^2+r_{+}^2) +\alpha \ln (\frac{r_{+}}{|\alpha| })
\right).
\end{equation}
\begin{figure}[h!]
\begin{center}
$%
\begin{array}{cc}
\includegraphics[width=80 mm]{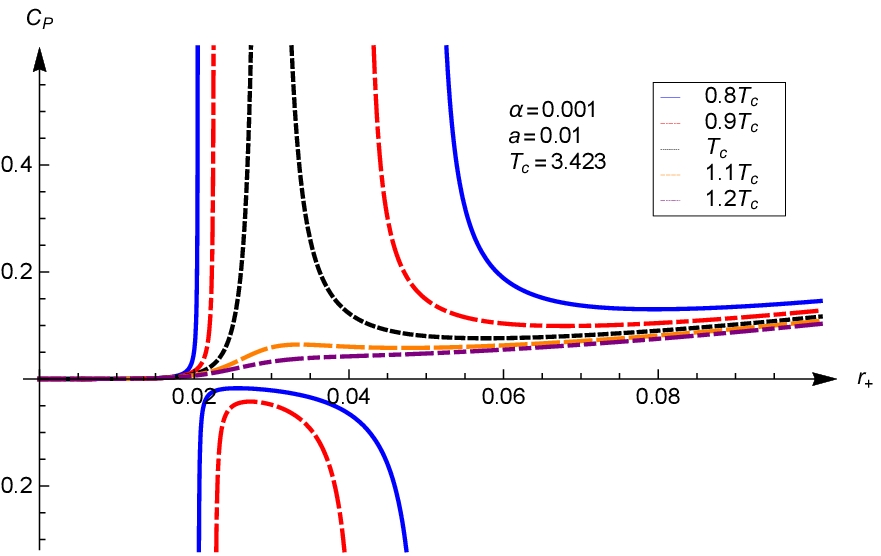} & %
\includegraphics[width=80 mm]{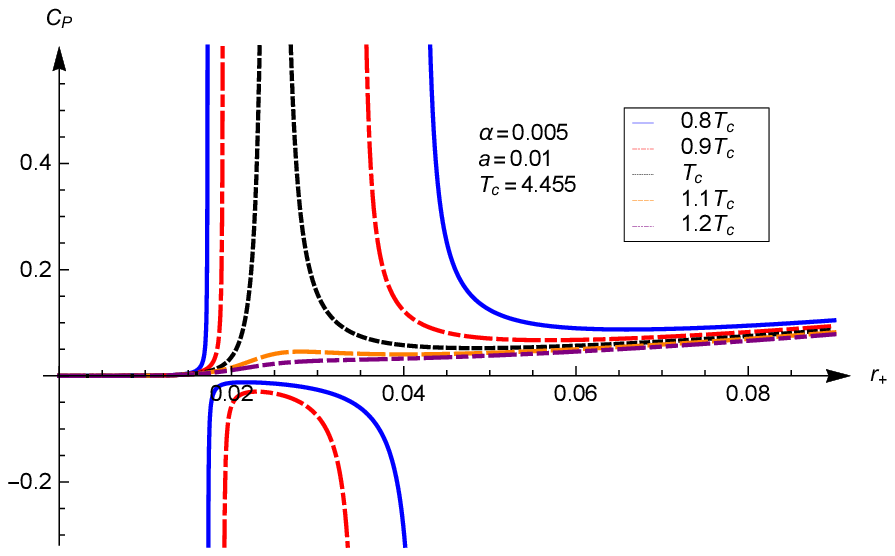}  %
\end{array}%
$%
\end{center}
\caption{Heat capacity versus $r_{+}$ for $a=0.01$ and
$\protect\alpha=0.001$ (left panel), and $\protect\alpha=0.005$
(right panel).} \label{FigC-Lambda}
\end{figure}
\begin{figure}[h!]
\begin{center}
$%
\begin{array}{cc}
\includegraphics[width=80 mm]{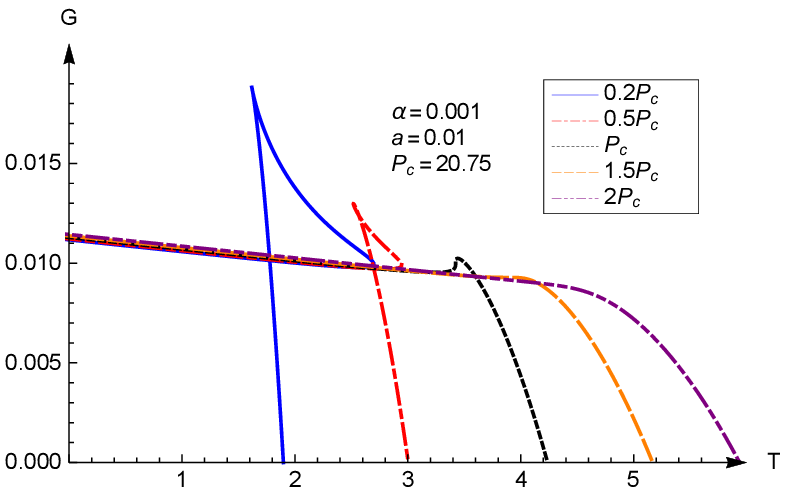} & %
\includegraphics[width=80 mm]{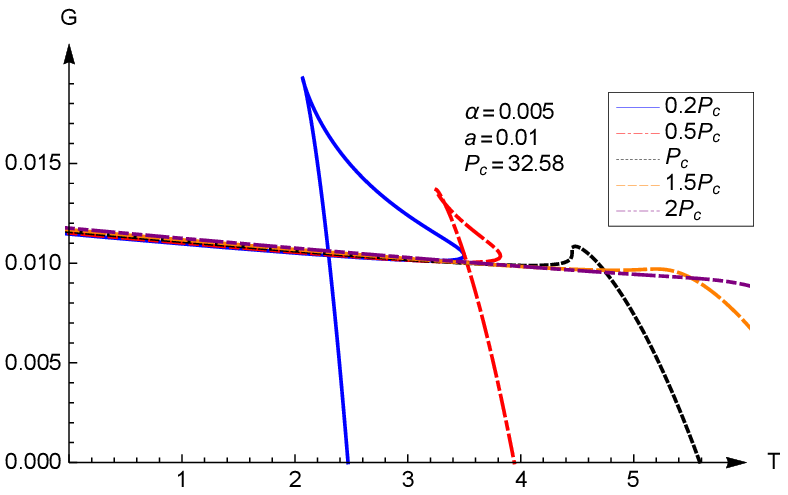}  %
\end{array}%
$%
\end{center}
\caption{Gibbs free energy versus $T$ for $a=0.01$ and
$\protect\alpha=0.001$ (left panel), and $\protect\alpha=0.005$
(right panel). } \label{FigG-Lambda}
\end{figure}
In Fig. \ref{FigC-Lambda}, we can find heat capacity behavior
versus critical radius  $ r_{+} $ for definite parameters. As we
expect from the first order phase transition, there are two
divergencies for pressures less than the critical pressure, which
are seen by red and blue colors. Negative heat capacity in this
region indicates unstable black hole, which is equivalent to the
oscillations of the isotherms under critical temperature in the $
P-r_{+} $ diagram (Fig. \ref{FigPV-Lambda}). These divergencies
are characteristics of the first order phase transition between
small and large black holes which are stable with positive heat
capacity. By changing the parameter $ \alpha $, a shift in the
horizontal axis occurs, which as mentioned before, increasing this
parameter results to decreasing of critical radius.

Gibbs free energy versus temperature is plotted in Fig.
\ref{FigG-Lambda}. The swallow-tail behavior for pressures less
than critical pressure, indicates a first order transition.
Besides, we see by increasing parameter $ \alpha $, critical
temperature increases and phase transition occurs as before.

Here, we are looking for the possible phase transition with
vanishing or constant $\Lambda $. To do so, we can define the
following ad hoc relation for the pressure
\begin{equation}
P=\frac{|\alpha |}{r_{+}^{3}}.  \label{newP}
\end{equation}

As one confirms, this pressure is related to the event horizon
radius which is the same as the pressure of black holes in dilaton
gravity. Inserting Eq. (\ref{newP}) into the relation of
temperature, one finds
\begin{equation}
P=\frac{4\pi T}{r_{+}}\left( 1+\frac{a^{2}}{r_{+}^{2}}\right) \left( 1+\frac{%
\Lambda a^{2}}{3}\right) +\frac{1}{r_{+}^{2}}\left( -1+\frac{%
\Lambda a^{2}}{3}\right) +\Lambda+\frac{a^2}{r_{+}^{4}} .  \label{newP2}
\end{equation}%
Having the equation of state at hand, we are in a position to
obtain the critical quantities via Eq. (\ref{dp-ddp}), as
\begin{eqnarray}
r_{c} &=&\frac{-a\sqrt{6\Gamma (\Gamma -\Theta -6)}}{2\Gamma },
\label{critical} \\
T_{c} &=&\frac{\sqrt{6}\Gamma ^{2}(\Gamma -\Theta -2)}{6\pi a(\Gamma
+6)(3\Gamma -\Theta -6)\sqrt{\Gamma \left( \Gamma -\Theta -6\right) }}, \\
%
P_{c} &=&\frac{2}{9 } \frac{\Gamma ^2}{a^2 (\Gamma -\Theta -6)}
\left[1+\frac{2}{(\Gamma -\Theta -6)} \left(1-\frac{
(\Gamma-\Theta -2) (5 \Gamma -3 \Theta -18)}{3 (3\Gamma -\Theta
-6)}\right)\right]+\Lambda,
\end{eqnarray}%
where $\Gamma =\Lambda a^{2}-3$ and $\Theta =\sqrt{\Lambda
^{2}a^{4}-26\Lambda a^{2}+105}$.

\begin{figure}[h!]
\begin{center}
$%
\begin{array}{cc}
\includegraphics[width=80 mm]{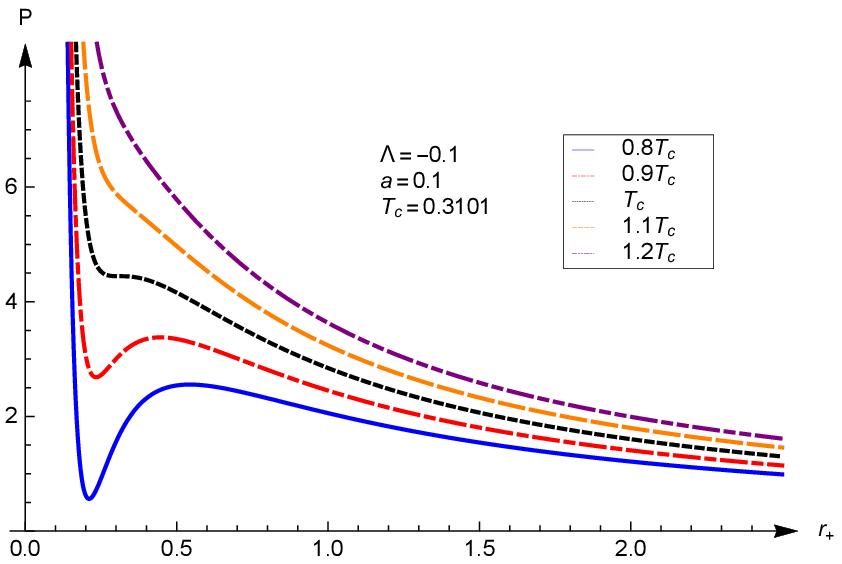} & %
\includegraphics[width=80 mm]{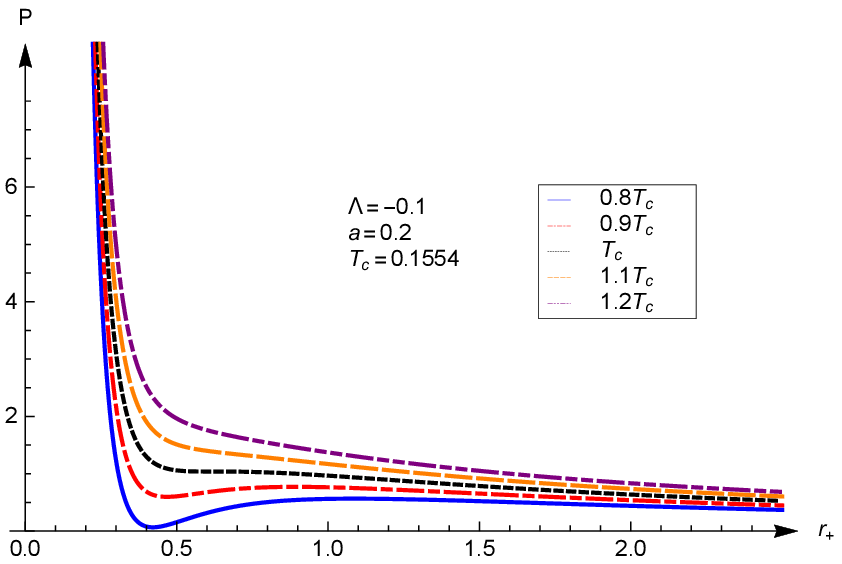}  %
\end{array}%
$%
\end{center}
\caption{Pressure versus $r_{+}$ for $\Lambda=-0.1$ and $a=0.1$
(left panel), and $a=0.2$  (right panel).} \label{FigPV-alpha}
\end{figure}

Now, we are going to investigate the possible phase transition
based on $P-r_{+}$ diagrams (see Fig. \ref{FigPV-alpha}).
According to this figure, we observe a van der Waals like behavior
which hints us for a first order phase transition. According to
table II, one can find the effects of rotation parameter and
cosmological constant. It is seen that unlike the rotation
parameter, the cosmological constant does not have a considerable
effect on the critical quantities. However, as we mentioned
before, increasing the rotation parameter leads to obtaining the
criticality easier.


\begin{center}
\begin{tabular}{|c|c|c|}
\hline
\begin{tabular}{cccc}
\hline\hline
\hspace{0.4cm}$a$\hspace{0.4cm} & \hspace{0.4cm}$r_{c}$ \hspace{0.4cm} &
\hspace{0.4cm} $T_{c}$\hspace{0.4cm} & \hspace{0.4cm} $P_{c}$\hspace{0.4cm}
\\ \hline\hline
$0.01$ & $0.03102$ & $3.098$ & $454.2$ \\ \hline
$0.05$ & $0.1551$ & $0.6198$ & $18.08$ \\ \hline
$0.10$ & $0.3102$ & $0.3101$ & $4.446$ \\ \hline
$0.20$ & $0.6202$ & $0.1554$ & $1.038$ \\ \hline
$0.30$ & $0.9297$ & $0.1040$ & $0.4072$ \\ \hline
\end{tabular}
& \ \ \ \ \ \  &
\begin{tabular}{cccc}
\hline\hline
\hspace{0.4cm}$\Lambda $\hspace{0.4cm} & \hspace{0.4cm}$r_{c}$ \hspace{0.4cm}
& \hspace{0.4cm} $T_{c}$\hspace{0.4cm} & \hspace{0.4cm} $P_{c}$\hspace{0.4cm}
\\ \hline
$-0.4$ & $0.31008$ & $0.31077$ & $4.1530$ \\ \hline
$-0.3$ & $0.31012$ & $0.31053$ & $4.2506$ \\ \hline
$-0.2$ & $0.31015$ & $0.31030$ & $4.3483$ \\ \hline
$-0.1$ & $0.31018$ & $0.31007$ & $4.4459$ \\ \hline
$0$ & $0.31022$ & $0.30983$ & $4.5435$
\end{tabular}
\\ \hline
\end{tabular}
\\[0pt]
\textbf{Table II: }Left table: critical values for $\Lambda =-0.1$
with different rotation
parameter.\\[0pt]
Right table: critical values for $a=0.1$ with different $\Lambda $.
\end{center}
\begin{figure}[h!]
\begin{center}
$%
\begin{array}{cc}
\includegraphics[width=60 mm]{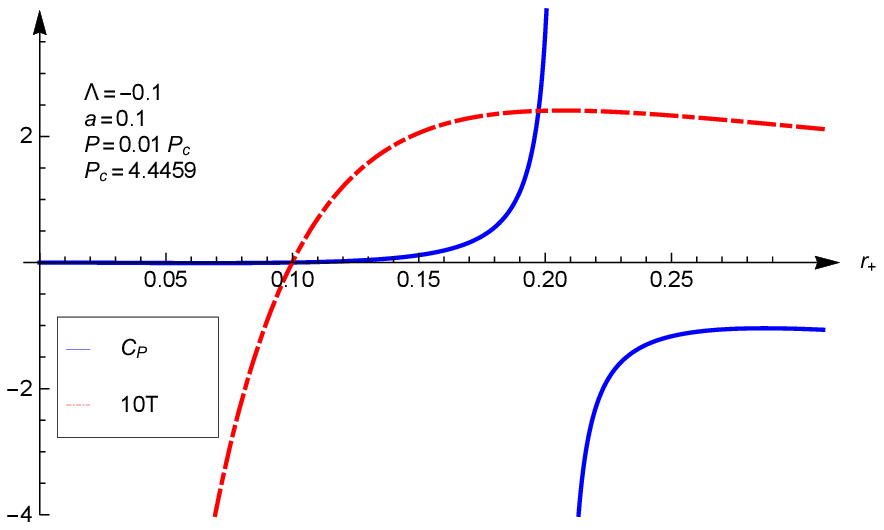} & %
\includegraphics[width=60 mm]{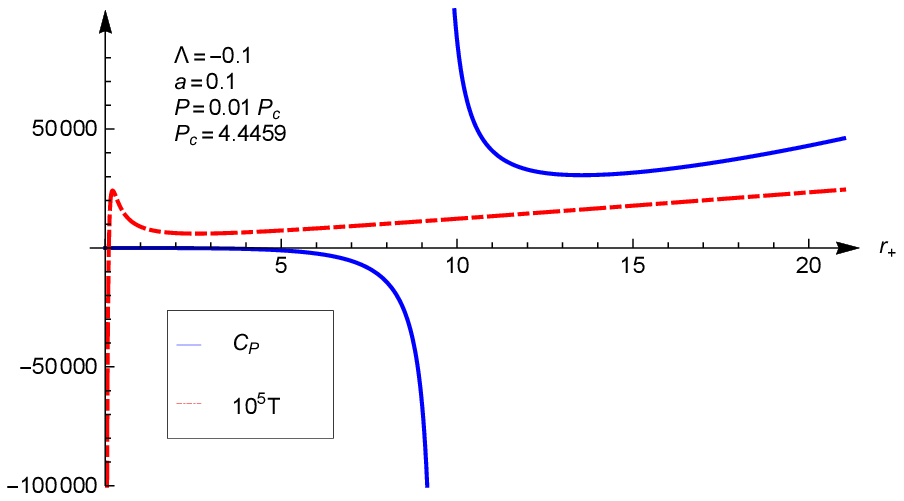}  %
\end{array}%
$%
\end{center}
\caption{Heat capacity versus $r_{+} $ for $ \Lambda =-0.1 $,
$a=0.1$  and $ P=0.01 P_{c } $, which shows divergencies in two
different regions. The dashed line corresponds to $ 10 T $ (right
panel) and $ 10^{5} T $(left panel).} \label{FigC-alpha}
\end{figure}

\begin{figure}[h!]
\begin{center}
$%
\begin{array}{cc}
\includegraphics[width=60 mm]{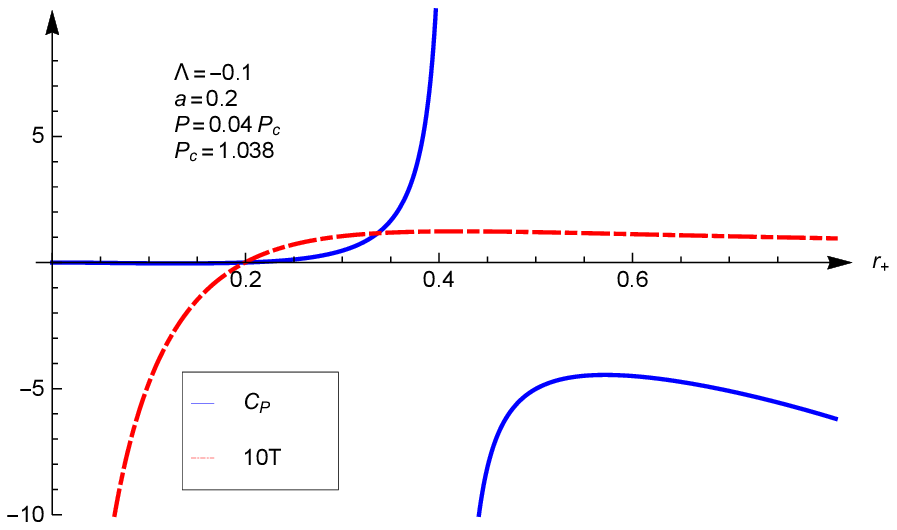} & %
\includegraphics[width=60 mm]{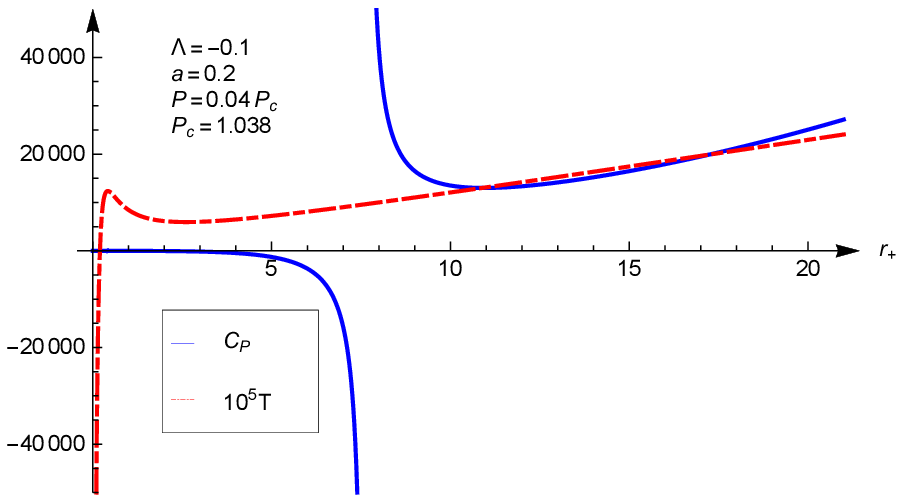}  %
\end{array}%
$%
\end{center}
\caption{Heat capacity versus $r_{+} $ for $ \Lambda =-0.1 $,
$a=0.2$  and $ P=0.04 P_{c } $, which shows divergencies in two
different regions. The dashed line corresponds to $ 10 T $ (right
panel) and $ 10^{5} T $(left panel).} \label{FigC-alpha1}
\end{figure}
\begin{figure}[h!]
\begin{center}
$%
\begin{array}{cc}
\includegraphics[width=80 mm]{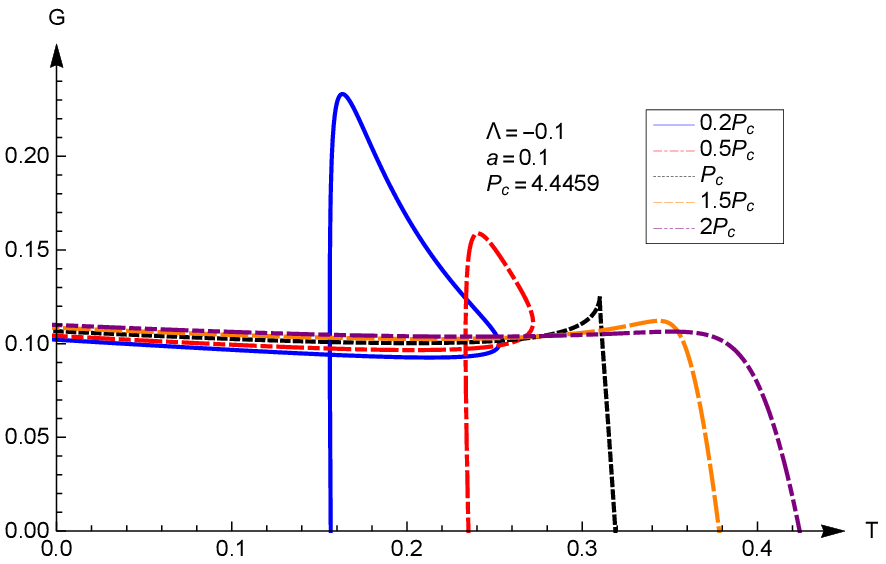} & %
\includegraphics[width=80 mm]{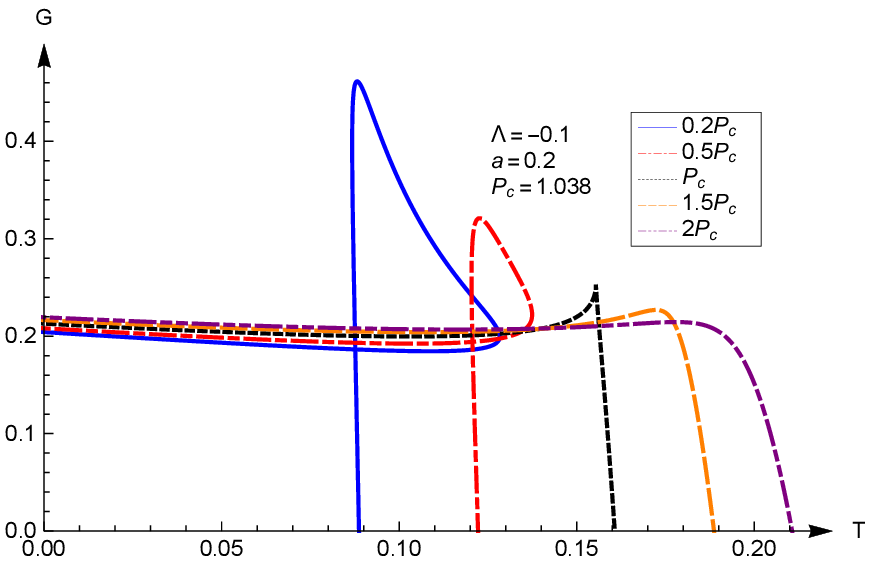}  %
\end{array}%
$%
\end{center}
\caption{Gibbs free energy versus $T$ for $ \Lambda =-0.1 $ and
$a=0.1$  (left panel), and $ \Lambda =-0.1 $ and $a=0.2$ (right
panel).} \label{FigG-alpha}
\end{figure}

Heat capacity and Gibbs free energy are given, respectively, as
\begin{equation}
C_{P}=\frac{6  \pi  r_{+}^2  (a^2+r_{+}^2) \left( (\Lambda -P)
r_{+}^2 +\frac{1}{3} \Lambda a^2 -1+\frac{a^2}{r_{+}^2}
\right)}{(3+\Lambda a^2) \left( (\Lambda +2 P)
r_{+}^4+(\frac{8}{3}\Lambda a^2+1) r_{+}^2+(\frac{\Lambda
a^2}{3}-4) a^2 - \frac{a^4}{r_{+}^2} \right)},
\end{equation}
\begin{equation}
G=\frac{(\Lambda -P) r_{+}^3 + \left(   \frac{\Lambda a^2}{3}-1
\right) +\frac{a^2}{r_{+}}}{4 \left(   \frac{\Lambda a^2}{3}+1
\right)^2  }+\frac{ r_{+}}{2} \left( 1+
\frac{a^2}{r_{+}^2}-\frac{1}{3} \Lambda  (a^2+r_{+}^2) -P r_{+}^2
\ln (|P| r_{+}^2)  \right).
\end{equation}

It is interesting to see the characteristic behaviors of the first
order phase transition by the ad hoc definition for pressure (Eq.
(\ref{newP})). Two divergencies for the heat capacity are shown
for different regions and parameters in Figs. \ref{FigC-alpha} and
\ref{FigC-alpha1}. The physical black holes with positive heat
capacity and temperature are seen just before the first divergency
and after the second one.

The swallow-tail like form of the Gibbs free energy for these two
figures is shown in Fig. \ref{FigG-alpha}, which assures existence
of the first order phase transition by defining pressure as Eq.
(\ref{newP}).


\section{Quasi-normal modes of a static PFDM black hole}

Let us consider the perturbation around the black hole background.
The black hole perturbations or quasinormal modes are a vital
source of gravitational waves generated as a result of a supernova
collapse. Numerous simulations of gravitational collapse of a
rotating massive star and merger of binary compact star system
result in the emission of gravitational waves which are directly
linked with quasinormal mode oscillations \cite{QNM2}. The process
of field oscillations in the curved background includes three
phases: initial perturbation, quasinormal mode oscillations and
the ensuing tails. The properties of oscillations just depend on
the second and third phases, so the study of QNM oscillations and
tails can help to understand the nature of black holes \cite{PRL}.
In the QNM oscillation phase, the function of field equation could
be written as $\Psi\sim e^{-i\omega t}R(r)Y(\theta,\varphi)$ where
$\omega$ is the eigenvalue whose real and imaginary parts
represent frequency and decay rate of QNM perturbations
respectively, so that the imaginary part should be negative for
stable black hole spacetime. The real part of the QNM frequency
denoted by $\omega_R$ determines the oscillation frequency while
the imaginary part $\omega_I$ determines the rate at which each
mode is damped as a result of emission of radiation.  With the
help of spherical harmonics, one can write $\Delta
Y(\theta,\varphi)=-L(L+1)Y(\theta,\varphi)$, where $L$ denotes
multipole quantum number. Since different black hole spacetimes in
different gravitational theories produce different QNMs frequency,
so investigating the details of QNMs is important and helpful to
discriminate different modified gravities or test the instability
of black hole solution. In the ensuing computations, we shall
ignore cosmological constant and the rotation for simplicity
reasons.

To study quasinormal modes, we focus on the following static
metric \cite{yang}
\begin{equation}
ds^2=-f(r)dt^2+\frac{dr^2}{f(r)}+r^2(d\theta^2+\sin^2\theta d\phi^2),
\end{equation}
where
\begin{equation}\label{blackeningFyang}
f(r)=1-\frac{2m}{r}+\frac{\alpha}{r}\ln\Big(\frac{r}{|\alpha|}\Big).
\end{equation}

\begin{figure}[h!]
\begin{center}
$%
\begin{array}{ccc}
\includegraphics[width=80 mm]{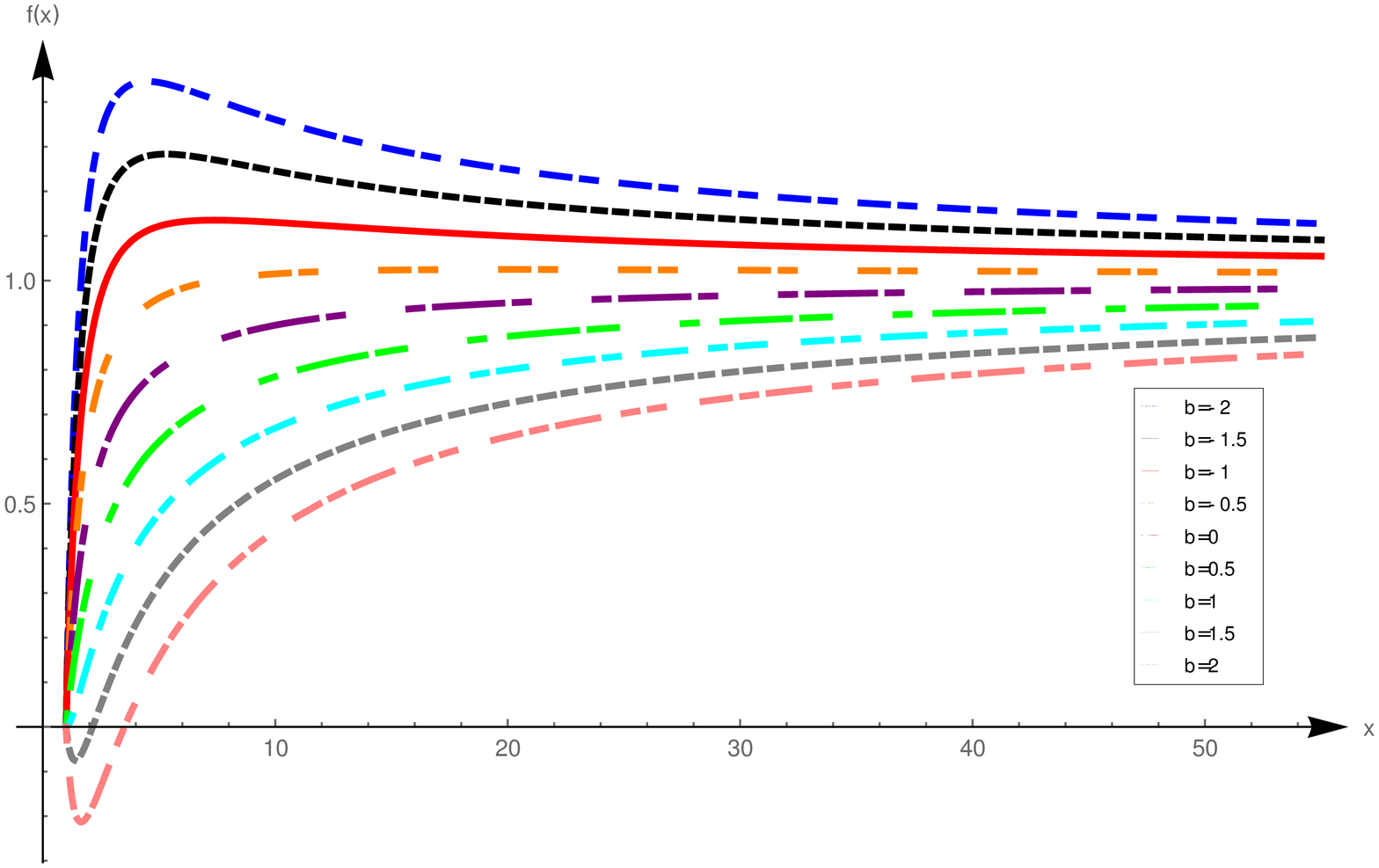}&\hspace{1cm} & \includegraphics[width=80 mm]{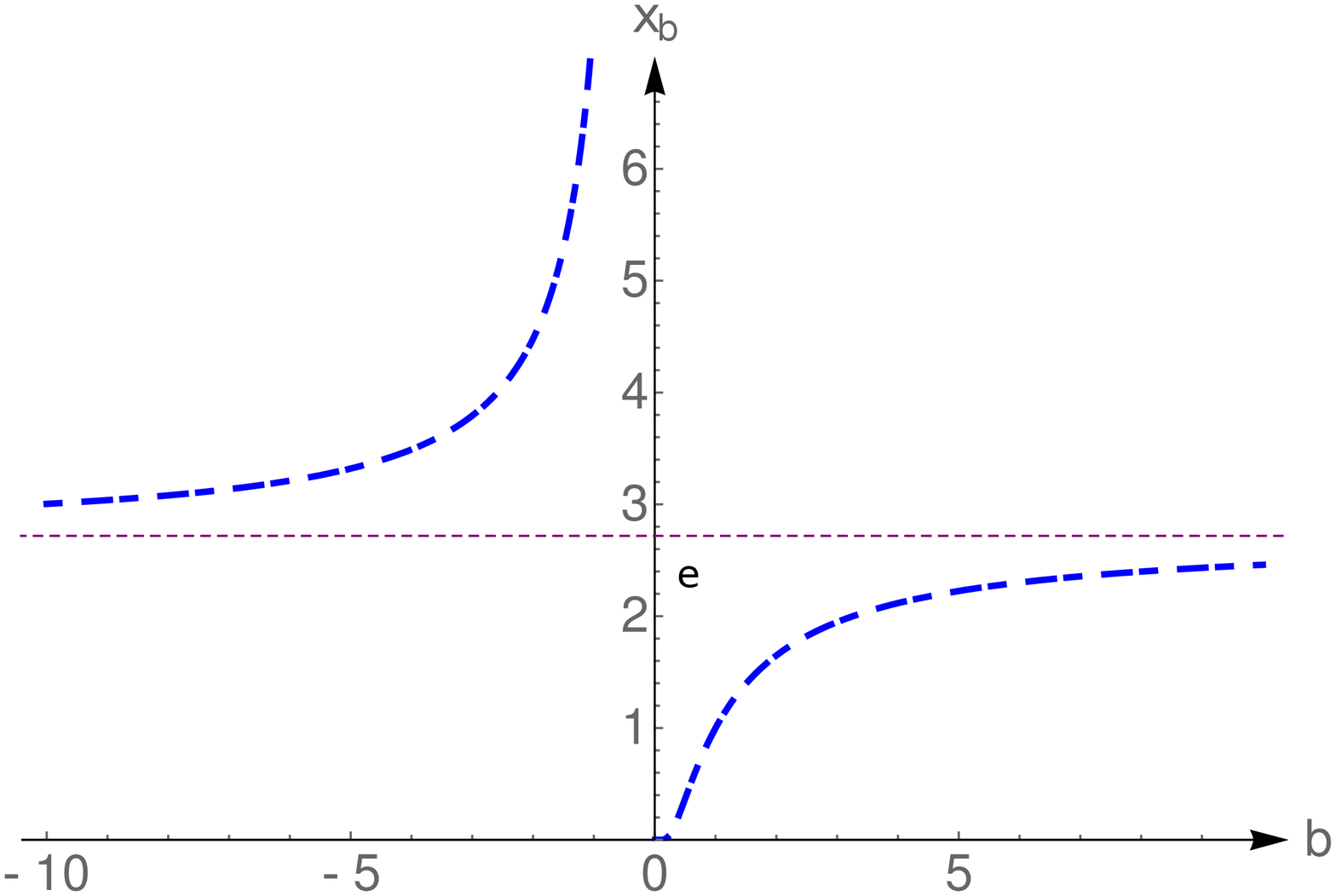}
\end{array}%
$%
\end{center}
\caption{The relation between $b$ and $f(x)$ (left) and the
relation between $b$ and $x_b$ (right).} \label{FIGxbLR}
\end{figure}

For computational and analytical simplicity, we employ the
following relation for the mass of black hole using  Eq. (\ref{blackeningFyang}) :
$2m=r_h+\alpha\ln\Big(\frac{r_h}{|\alpha|}\Big)$, which allows
rewriting the above equation as
\begin{equation}
f(r)=1-\frac{r_h}{r}+\frac{\alpha}{r}\ln\Big(\frac{r}{|\alpha|}\Big)-\frac{\alpha}{r}
\ln\Big(\frac{r_h}{|\alpha|}\Big)=1-\frac{1}{x}-\frac{b}{x}\ln(x),
\end{equation}
and entails $f(r_h)=0$, where $r_h=r_{+}$ denotes the size of
horizon radius, and $x=r/r_h$, $\alpha=-b\;r_h$. Interestingly,
$r_h$ is event horizon as $b<1$, while it becomes inner horizon as
$b>1$, see left panel of Fig. \ref{FIGxbLR}. As $b=0$, the black
hole becomes the Schwarzschild black hole, and the temperature of
black hole vanishes as $b=1$. As $b>0$, the position
$x_b=e^{1-1/b}$ which satisfies $f'(x_b)=0$ is less than the event
horizon, but as $b<0$, $x_b$ is larger than the event horizon so
that a maximum value exists outside the event horizon, see right
panel of Fig. \ref{FIGxbLR}. We find the form of $f(r)$ with a
maximum point which is similar to non-Schwarzschild solution in
quadratic and higher derivative gravity \cite{k1a,k1b,k1c}.

We consider a massless scalar perturbation in the background of
the black hole spacetime. The equation of motion of a minimally
coupled scalar field is given by:
\begin{equation}
\nabla^\mu\nabla_\mu \Phi=0.
\end{equation}
We express the scalar field in the form of separation of variables
i.e.
$\Phi(x^\mu)=\sum\frac{\Psi_L(r)}{r}Y_{Lm}(\theta,\phi)e^{-i\omega
t}$. Since we are dealing with spherically symmetric spacetimes
the solution will be independent of m, thus this subscript can be
omitted. Introducing the tortoise coordinate
$r_*=\int\frac{dr}{f(r)},$ having range between
$(-\infty,+\infty)$, it is possible to write the radial part of
the wave equation (also known as Regge-Wheeler form) as follows
\begin{equation}
\label{QNMsEqu}
\frac{d^2\Psi}{dr_*^2}+[\omega^2-V(r)]\Psi(r_*)=0,
\end{equation}
where the effective potential is given as
\begin{equation}
V(r)=f(r)\Big( \frac{f'(r)}{r}+\frac{L(L+1)}{r^2} \Big).
\end{equation}

\begin{figure}[h!]
\begin{center}
$
\begin{array}{cc}
\includegraphics[width=60 mm]{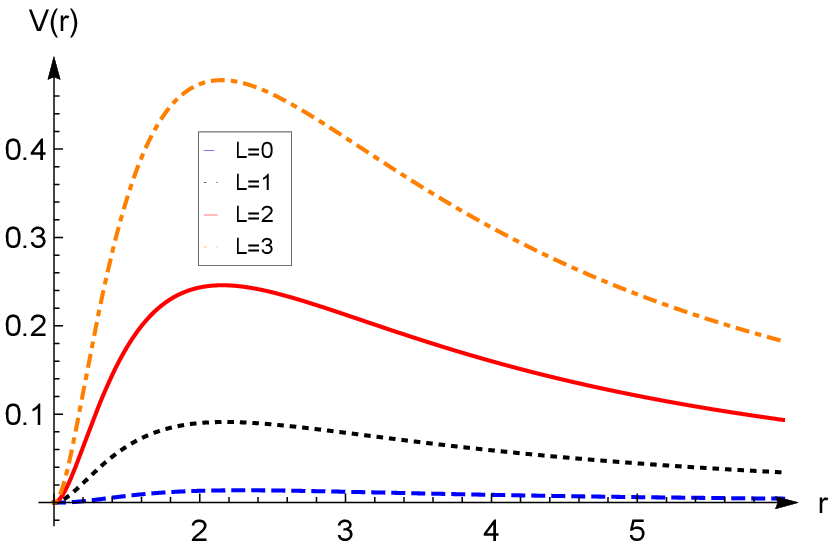} & \includegraphics[width=60
mm]{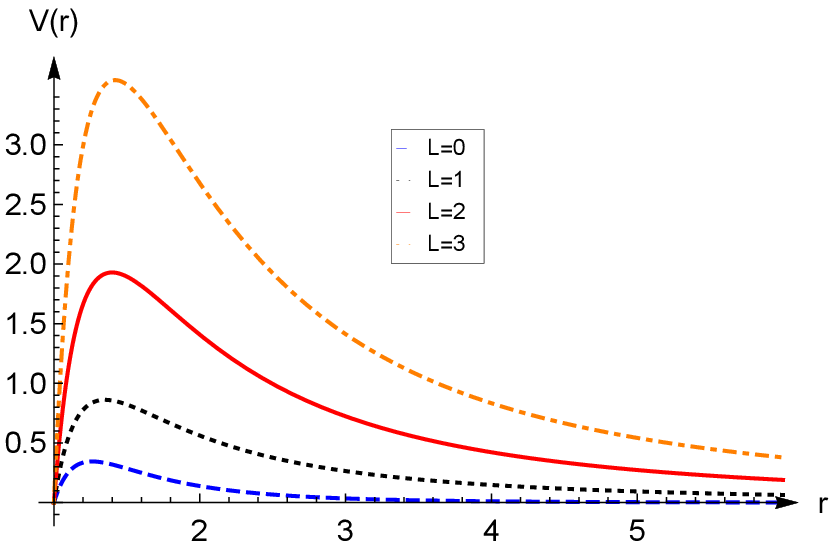} \\
\includegraphics[width=60 mm]{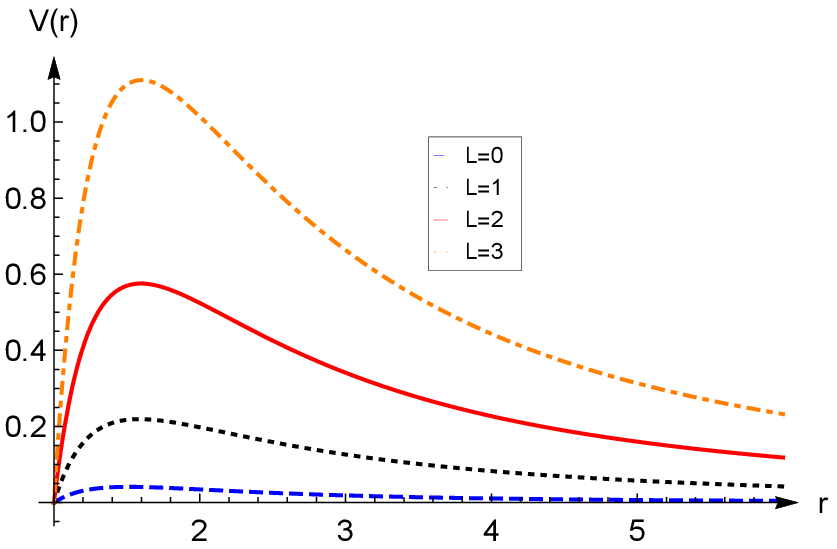}  &  \includegraphics[width=60
mm]{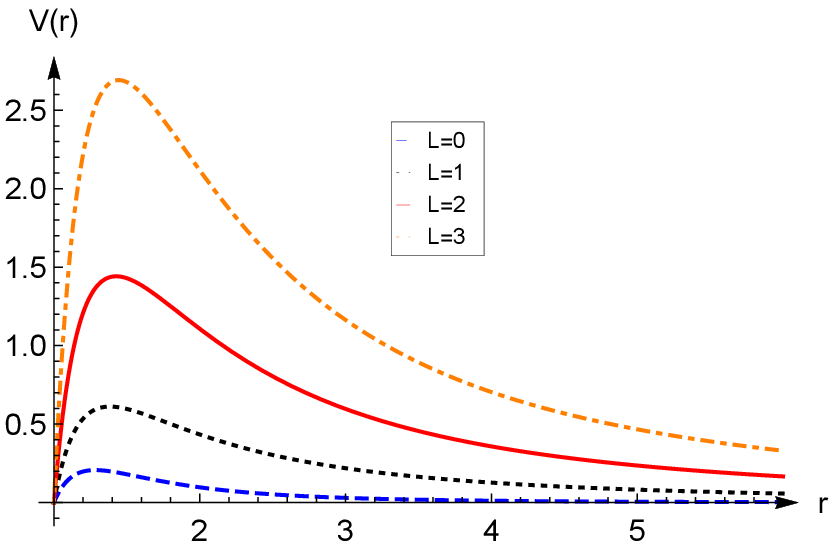}
\end{array}
$
\includegraphics[width=70 mm]{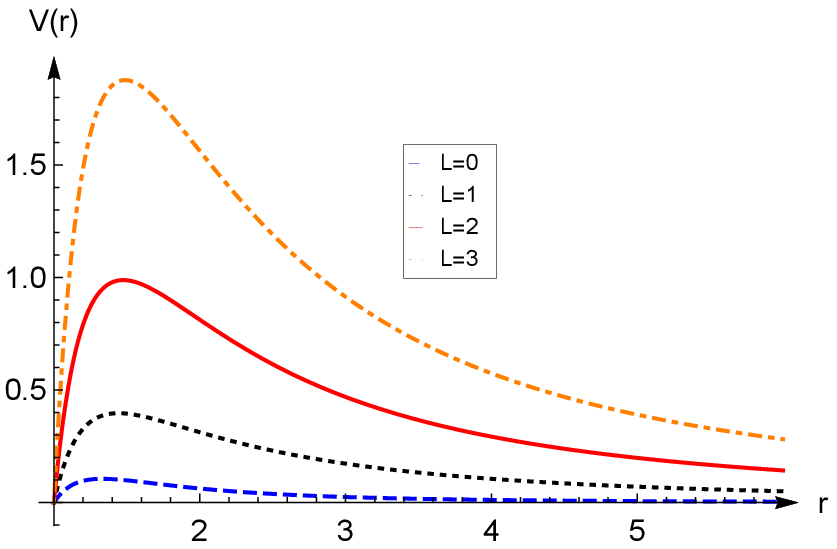}
\end{center}
\caption{$V(r)$ versus $r$ for $b=1$ (up-left panel), $b=-1$
(up-right panel), $b=0.5$ (middle-left panel), $b=-0.5$
(middle-right panel) and $b=0$ (down panel).  } \label{Vdiff-b}
\end{figure}


\begin{figure}[h!]
\begin{center}
$%
\begin{array}{cc}
\includegraphics[width=60 mm]{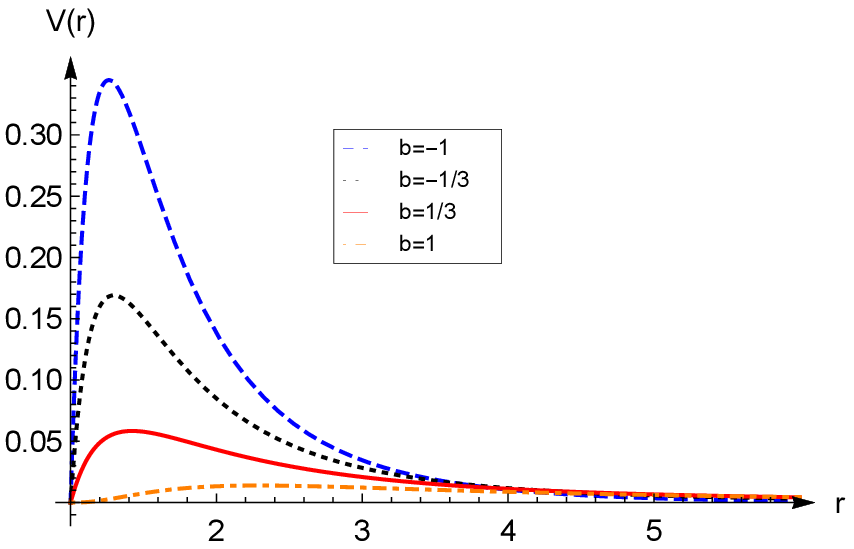} & \includegraphics[width=60
mm]{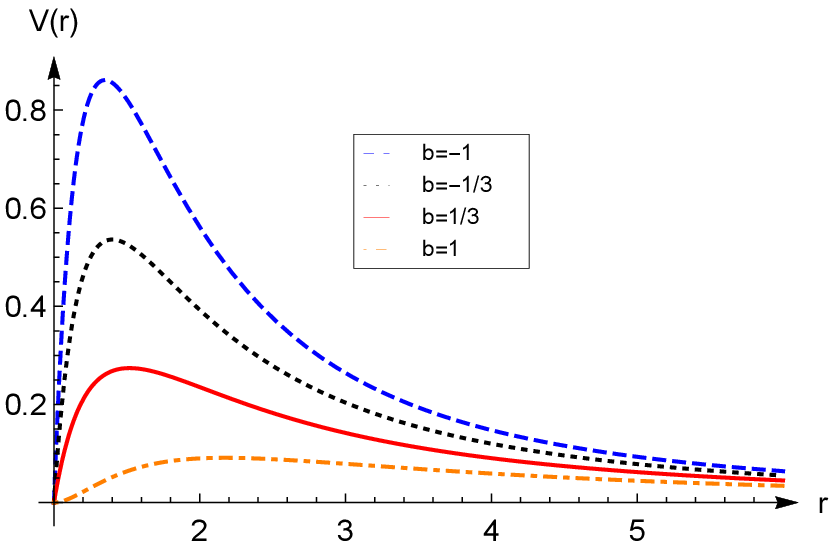} \\
\includegraphics[width=60 mm]{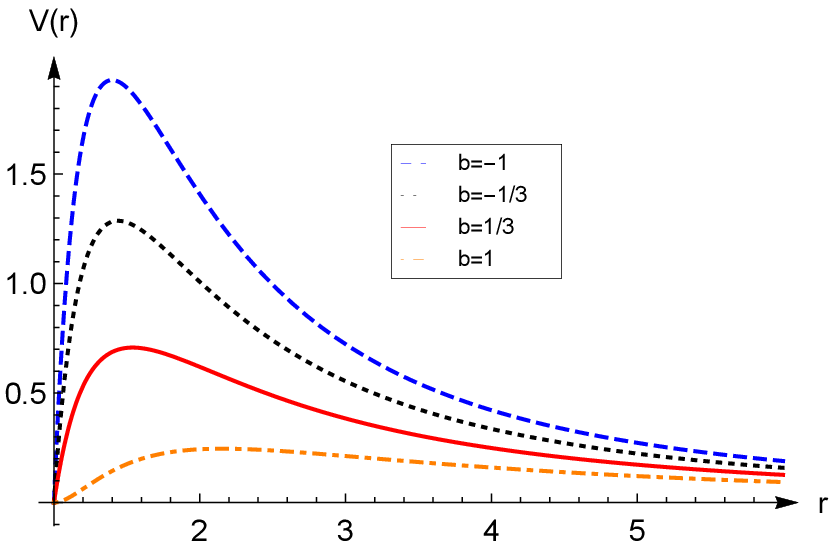} & \includegraphics[width=60
mm]{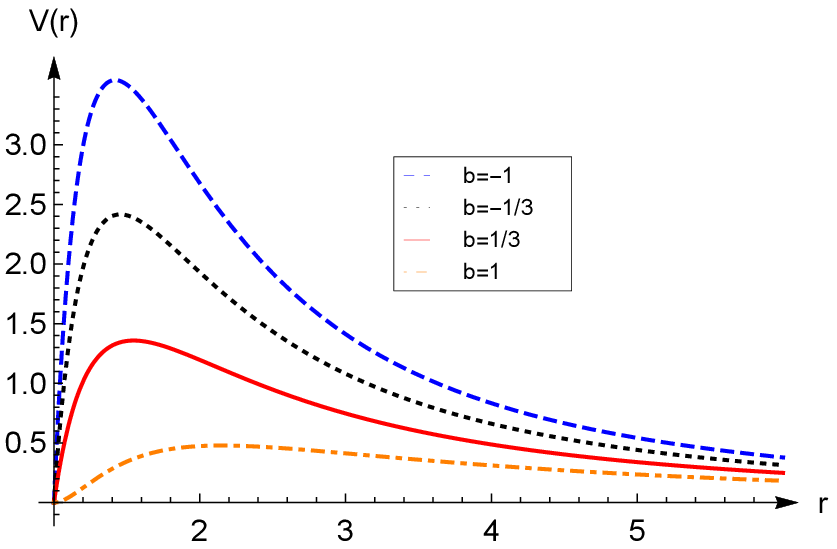}
\end{array}%
$%
\end{center}
\caption{$V(r)$ versus $r$ for $L=0$ (up-left panel), $L=1$
(up-right panel), $L=2$ (down-left panel) and $L=3$ (down-right
panel).} \label{Vdiff-L}
\end{figure}


There are numerous numerical approaches for determining the
solution, known as quasinormal modes with complex frequencies,  of
the above wave equation (see \cite{QNM4}). However, we shall use
the sixth order Wentzel-Kramers-Brillouin (WKB) approximation
method to calculate the quasinormal modes. The WKB method was
originally proposed by Schutz, and Iyer and Will at third order,
independently \cite{will1,will2} while later was extended to sixth
order in \cite{kono}. This approach is useful to obtain the QNMs
for a full range of parameters and has the best accuracy for
$L\geq n$. In general, for a given effective potential $V(r)$, the
sixth order WKB formula has the form
$$\frac{i(\omega^2-V_0)}{\sqrt{-2V_0''}}-\sum_{i=2}^{6}\Lambda_i=n+\frac{1}{2},$$
where $V_0$ denotes the numerical value of the effective potential
evaluated at the peak (or the maximum) while similarly $V_0''$
denotes the corresponding second derivative evaluated at the
maximum. Also, the correction terms $\Lambda_i$'s are
corresponding to the the i-th order of WKB method depending on the
value of the effective potential, $V(r)$ and its derivatives at
the local maximum \cite{will2,kono}, and $n$ is the overtone
number.

Considering the functional form of $V(r)$ and Figs. \ref{Vdiff-b}
and \ref{Vdiff-L}, we find that the effective potential is
constant at the boundaries (the event horizon and the infinity),
and it rises to a maximum at an intermediate point $r = r_{0}$.
Acceptable accuracy of the (sixth order) WKB method is guaranteed
since such a method is based on the matching of WKB expansion of
the wave function at the boundaries with the Taylor expansion near
the local maximum of the potential barrier through the two turning
points. In other words, WKB method can be used for an effective
potential that forms a potential barrier and takes constant values
at the boundaries (See Figs. \ref{Vdiff-b} and \ref{Vdiff-L} for
more details regarding the effect of different parameters on the
peak of the potential).

\begin{figure}[h!]
\begin{center}
$%
\begin{array}{cc}
\includegraphics[width=80 mm]{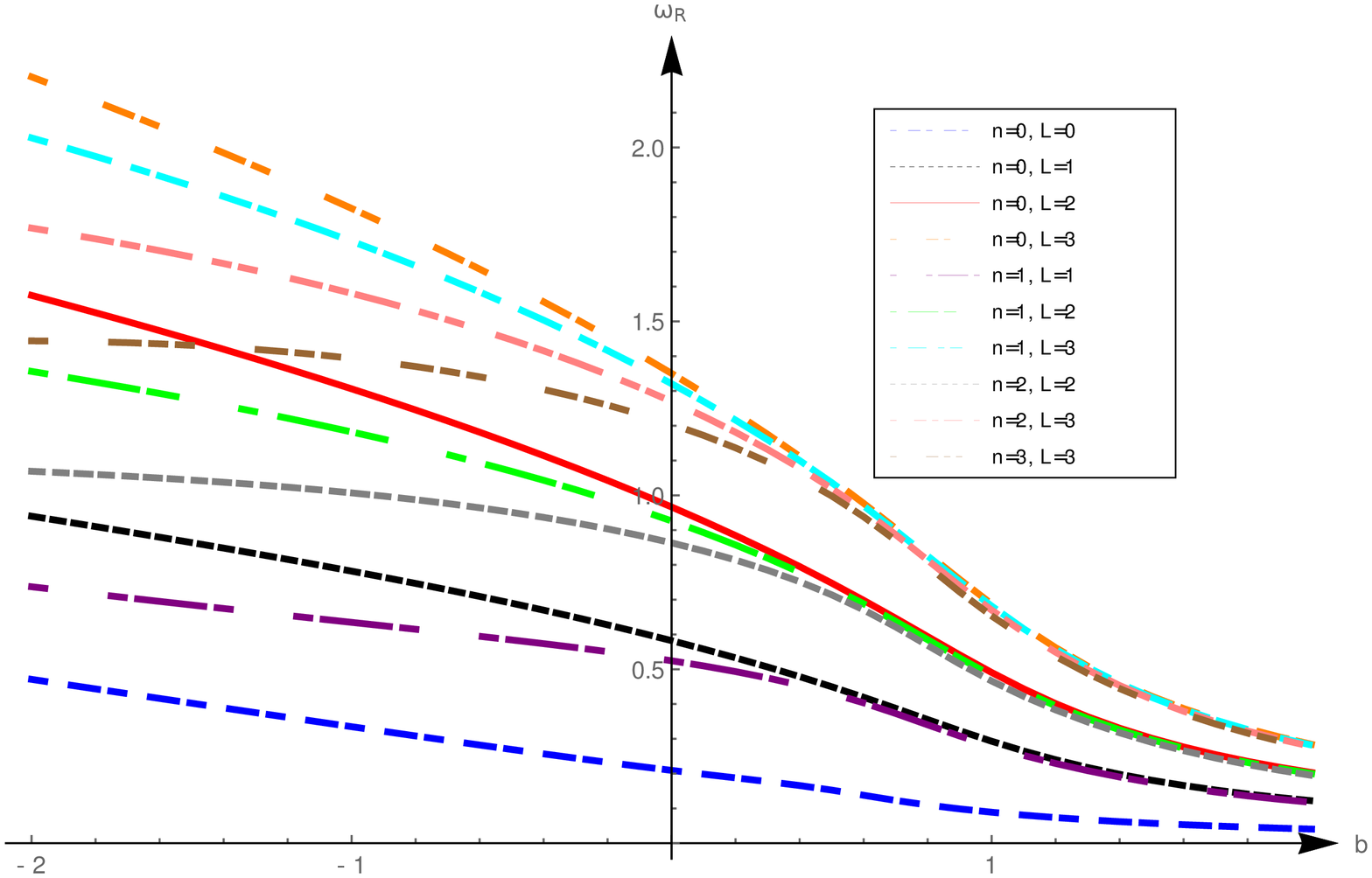} & %
\includegraphics[width=80 mm]{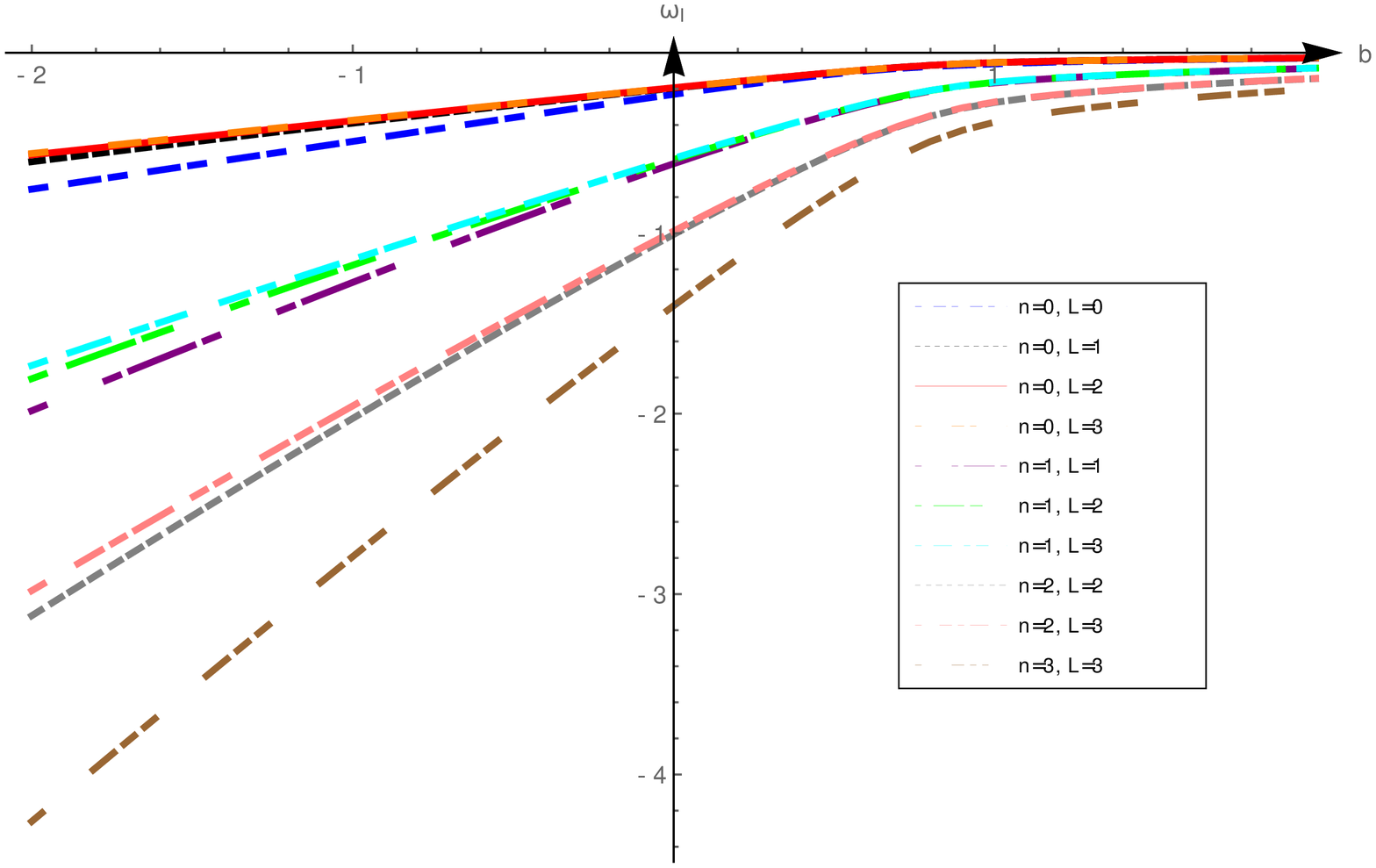}%
\end{array}%
$%
\end{center}
\caption{Scalar quasinormal modes of the black hole for different values of angular (or multipole) quantum number $L$ and overtone $n$.} \label{WKBQNMs}
\end{figure}

From the Figs. \ref{WKBQNMs}, we have plotted the real and
imaginary components of oscillation frequency against parameter
$b$ and shown that QNMs perturbation will increase but the damping
of QNMs is faster as $b$ reduces. Further, $\omega_I<0$ depicting
that black hole configuration is stable after perturbation.

\begin{figure}[h!]
\begin{center}
$%
\begin{array}{cc}
\includegraphics[width=80 mm]{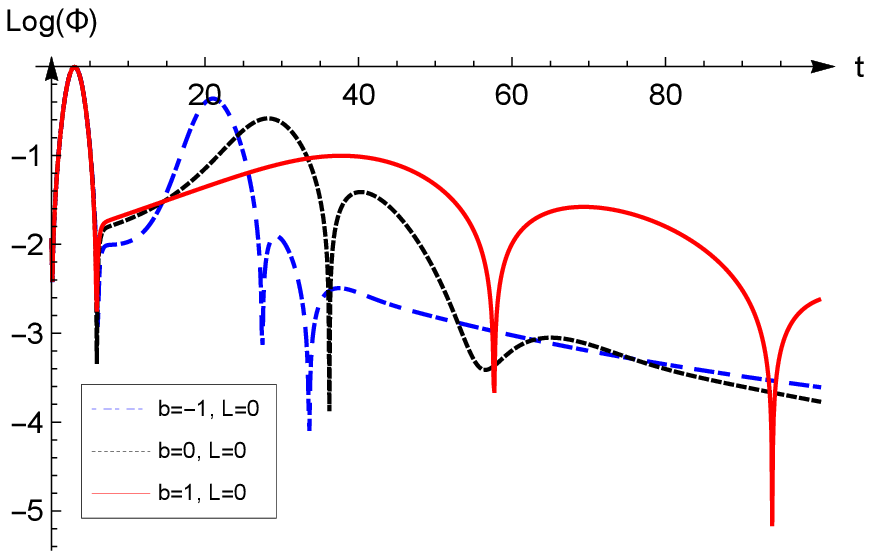}\includegraphics[width=80 mm]{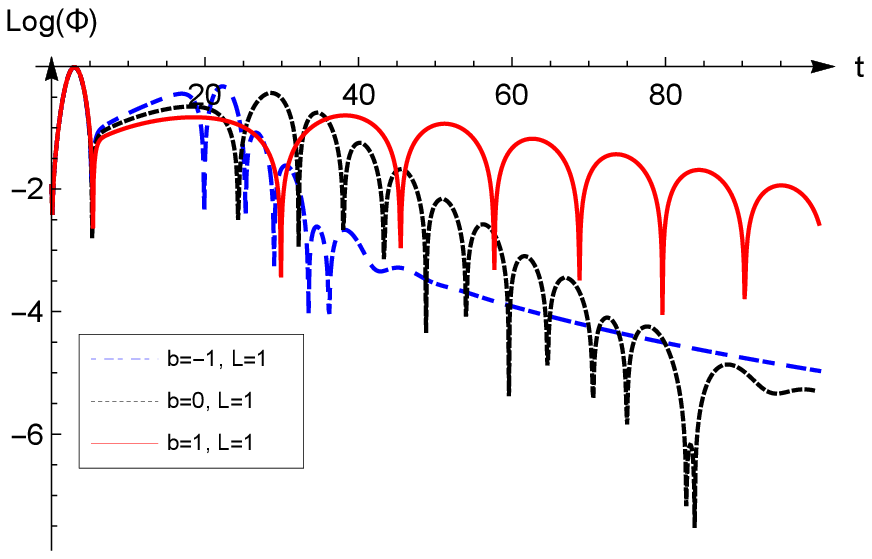}\\
\includegraphics[width=80 mm]{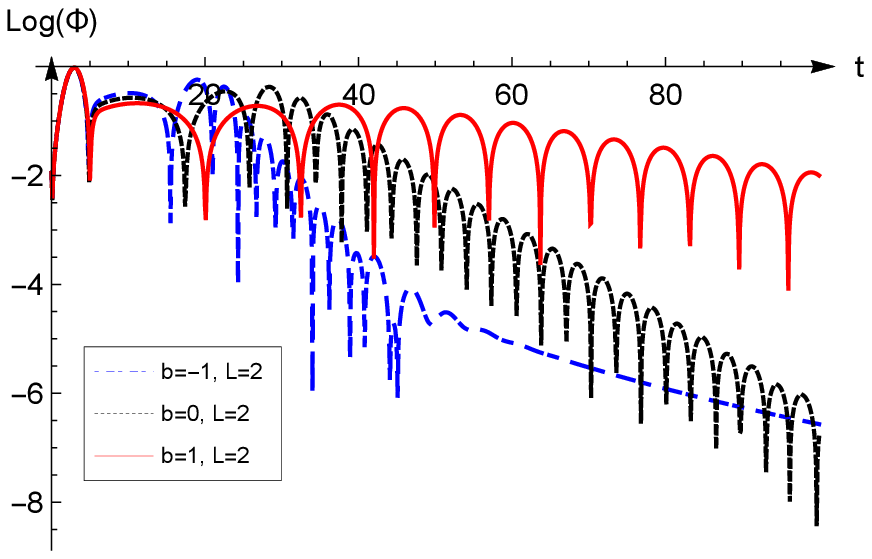}\includegraphics[width=80 mm]{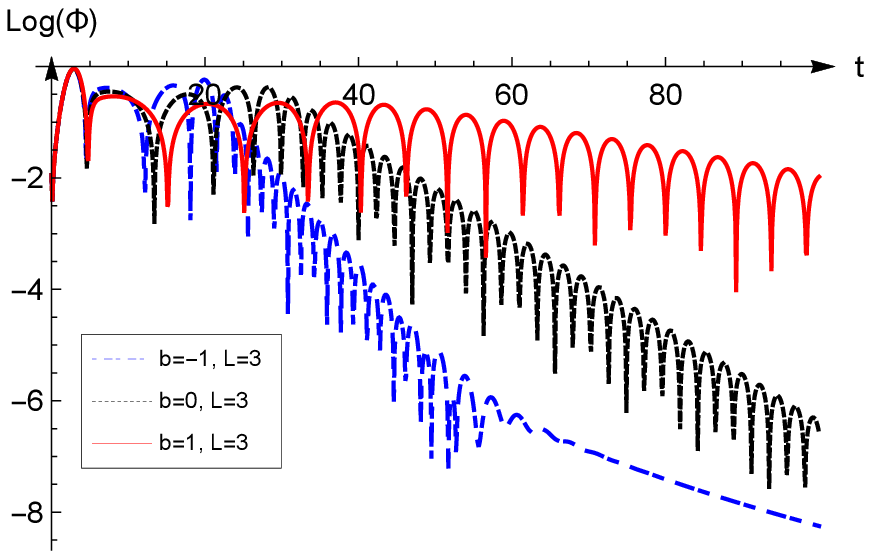}
\end{array}%
$%
\end{center}
\caption{Dynamical evolution  of a scalar field perturbation to
the black hole.} \label{FDMQNMs}
\end{figure}

\begin{figure}[h!]
\begin{center}
\includegraphics[width=140 mm]{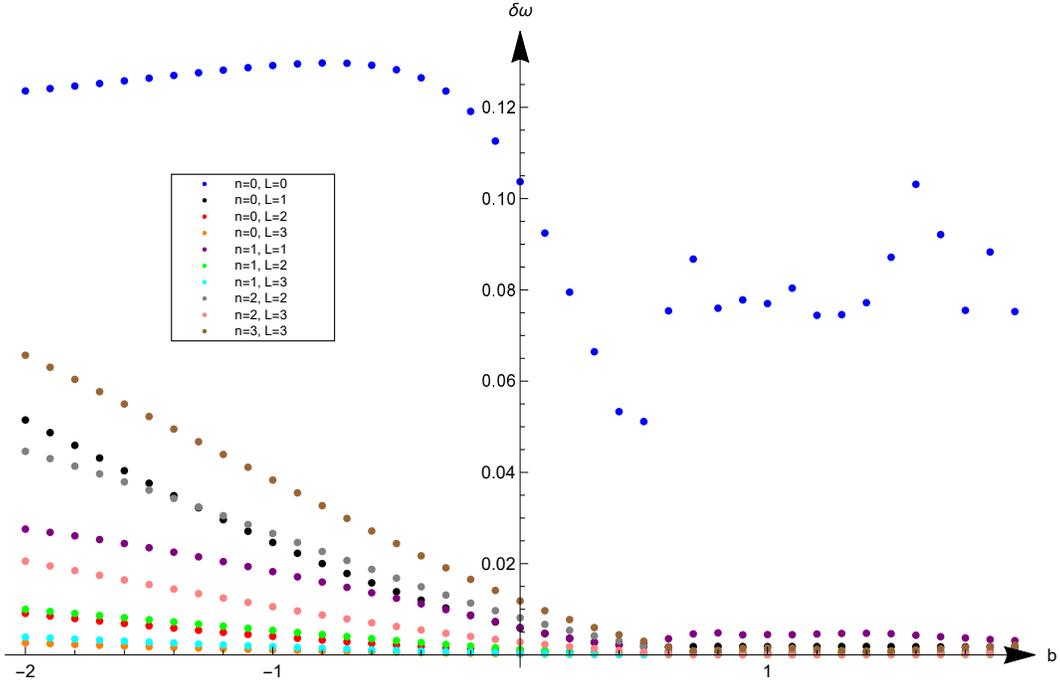}
\end{center}
\caption{A plot showing relative errors between WKB methods of
order 6 and 3.} \label{error}
\end{figure}

In order to study the dynamical properties of QNMs perturbation,
we show the details of black hole oscillations by using finite
difference method following \cite{jin}. Firstly, we rewrite
Eq.(\ref{QNMsEqu}) as
\begin{equation}
\frac{\partial^2\Phi}{\partial
r_*^2}-\frac{\partial^2\Phi}{\partial t^2}=V(r)\Phi.
\end{equation}
After the transformation $u=t-r_*$ and $v=t+r_*$, above equation becomes
\begin{equation}
\frac{\partial^2\Phi}{\partial u \partial
v}+\frac{1}{4}V(r)\Phi=0.
\end{equation}

Therefore, we can use finite difference method to solve above
equation and show the process of black hole oscillations in Fig.
\ref{FDMQNMs}. The oscillations are damped for fixed $L$ and three
values of $b$. In each figure, it is observed that $b=-1$ case has
the fastest damping ending up in a tail, followed by $b=0$ which
has a rather slow damping speed with an observed tail while $b=1$
has the slowed decaying rate. In Fig. \ref{error}, we have plotted
relative errors between the WKB methods of order $3$ and $6$, by
defining a parameter $$\delta \omega\equiv
|\frac{\omega_6-\omega_3}{\omega_6}|,$$ where $\omega_3$ and
$\omega_6$ are the values of QNM frequencies obtained by the WKB
method of orders $3$ and $6$, respectively. For different values
of parameters $n$ and $L$, the relative error in absolute form
$\delta\omega$ increases along negative $b-$axis, while along
positive $b-$axis the relative error $\delta\omega$ decreases and
both orders of WKB method give same kind of approximation to QNM
frequencies. In summary, the errors are very small with order less
than $0.1$, hence the numerical analysis is considered to be
credible.

\section{Conclusion}

In this paper, we considered the Kerr-AdS black hole surrounded by
the perfect fluid dark matter (PFDM). We have done a detailed
analysis of the effect of PFDM on the thermodynamical behavior of
this kind of black holes. By giving dynamics to the cosmological
constant and it's association with pressure i.e. $
P=\frac{-\Lambda}{8 \pi} $, we worked in the extended phase space
to study the phase behavior. We found the van der Waals like
behavior for the solution and by a numerical method, we observed
that increasing the rotation parameter (decreasing $\alpha$) leads
to increasing the critical horizon radius and decreasing the
critical temperature and pressure. Two divergencies in the $
C_P-r_+ $ plots and the swallow-tail shape in the $ G-T $ plots
for pressures less than the critical pressure indicate that these
black holes enjoy the first order phase transition.

On the other hand, by dimensional analysis, we proposed another
relation for the pressure $ P=\frac{|\alpha |}{r_{+}^{3}} $. In
this case, pressure is related to the PFDM parameter and the
horizon radius. Analytic relations for the critical quantities are
found and the results show that unlike the rotation parameter, the
cosmological constant does not have a considerable effect on the
critical quantities. However, as before, increasing the rotation
parameter leads to obtaining the criticality easier.
Interestingly, we observe that this ad hoc definition for the
pressure also results in the first order phase transition for the
introduced black hole solutions.

In the last part, we studied the QNMs for the static metric in the
presence of the PFDM. Using the sixth order WKB method, we
investigated the massless scalar quasinormal modes (QNMs) for the
static spherically symmetric black hole surrounded by dark matter.
Using the finite difference scheme, the dynamical evolution of the
QNMs are also discussed for different values of angular momentum
and overtone parameters. The QNM oscillations are damped in all
cases of considered $L$ and $b$, however for $b=-1,0$, the damping
is faster as compared to $b=1$. Here we ignored the effects of
cosmological constant and spin while studying the QNMs, however
one may employ the method of Horowitz-Hubeny \cite{hh} to analyze
QNM frequencies for static black holes with AdS boundary. This is
left as future work.

Since the spectrum of gravitational QNMs perturbations can be
traced by gravitational wave detectors, it will be interesting to
study the response of the solutions under tensor perturbation.
This work may be addressed in independent work.

\section*{Acknowledgement}

We are grateful to the anonymous referee for the insightful
comments and suggestions, which have allowed us to improve this
paper significantly. SHH and AN wish to thank Shiraz University
Research Council. The work of SHH has been supported financially
by the Research Institute for Astronomy and Astrophysics of
Maragha, Iran.

\end{document}